\documentstyle[preprint,aps,prb]{revtex}
\begin{document}
\title{Exact solution of the Dirac equation  for  a Coulomb and a scalar
Potential in the presence of an  Aharonov-Bohm and a magnetic monopole fields}

\author{ V\'{\i}ctor M. Villalba\footnote{e-mail address:
villalba@dino.conicit.ve}}

\address{Centre de Physique Th\'eorique C.N.R.S. \\ Luminy- Case 907-F-13288
Marseille Cedex 9, France}

\address{Centro de F\'{\i}sica, Instituto Venezolano de Investigaciones
Cient\'{\i}ficas \\ IVIC, Apdo. 21827, Caracas 1020-A Venezuela}

\maketitle

\begin{abstract}
In the present article we analyze the problem of a relativistic Dirac electron
in the presence
of a combination of a Coulomb field, a $1/r$ scalar potential as well as a
Dirac magnetic
monopole and an Aharonov-Bohm potential. Using the algebraic method of
separation
of variables, the Dirac equation expressed in the local rotating diagonal gauge
is
completely separated in spherical coordinates, and exact solutions are
obtained.
We compute the energy spectrum and analyze how it
depends on the intensity of the Aharonov-Bohm and the magnetic monopole
strengths.
\end{abstract}

\section{Introduction}

The Dirac equation is a system of four coupled partial differential which
describes the relativistic electron and other spin 1/2 particles. Despite
the remarkable effort made during the last decades in order to find exact
solutions for the relativistic Dirac electron the amount of solvable
configurations is relatively scarce, being the Coulomb problem perhaps the
most representative example and also one of the most discussed
and analyzed problems in relativistic quantum mechanics. Among the different
approaches available in the literature for discussing the Dirac-Coulomb
problem in the presence of other interactions like the Aharonov-Bohm field
or any other electromagnetic potential we have the quaternionic approach
proposed by Hautot \cite{Hautot}, the St\"ackel separation method developed
by Bagrov{\it \ et al }\cite{Bagrov}, the algebraic method of separation of
variables \cite{alg1,alg2}, the shift operator method \cite{Lange}, and the
algebraic method proposed by Komarov and Romanova \cite{Komarov1}.

Recently, Lee Van Hoang{\it \ et al }\cite{Komarov} have solved the
Dirac-Coulomb problem when an Aharanov-Bohm and a Dirac magnetic monopole
fields are present. The authors use, for tackling the problem, a two
dimensional complex space which results after applying the
Kustaanheimo-Stiefel transformation \cite{Kust} on the three space
variables, reducing in this way the Kepler problem to an oscillator problem.
This idea lies on the utilization of a SU(2) dynamical algebra for computing
the resulting energy spectrum \cite{Kibler}, which, like the spinor
solution, is expressed in terms of intrinsic coordinates appearing after
using the complex space. The utilization of different techniques for
studying the Dirac-Coulomb field in the presence of an Aharonov-Bohm field or
a magnetic monopole could give rise to the idea that this problem is not
soluble without introducing new variables or additional conserved
quantities. Here it is shown that using the algebraic method of separation
of variables it is possible to solve the Dirac equation in the presence of a
Coulomb field and a scalar 1/r potential with an Aharonov-Bohm and a
magnetic monopole fields. The advantage of this approach is that does not
require the introduction of non bijective quadratic transformations, also it
becomes clear the role played by the Dirac magnetic monopole as well as the
Aharonov-Bohm field in the angular dependence of the spinor $\Psi (\vec r)$
solution of the Dirac equation.

The article is structured as follows: In Sec. II, applying a pairwise scheme
of separation, we separate variables in the Dirac equation expressed in the
local rotating frame, we separate the radial dependence from the angular
one. In Sec. III. the angular dependence is solved in
terms of Jacobi Polynomials. In Sec IV, the separated radial equation is
solved and the energy spectrum is computed. In Sec. V, we discuss the
influence of the Aharonov-Bohm field and the magnetic monopole on the energy
spectrum.

\section{Separation of variables}
In this section we proceed to separate variables in the Dirac equation when
a Coulomb field, a scalar 1/r potential as well as a Dirac magnetic monopole
and a Aharonov-Bohm field are present. For this purpose, we write in
spherical coordinates the covariant generalization of the Dirac equation
\begin{equation}
\label{Dirac}\left\{ {\tilde \gamma }^\mu (\partial _\mu -\Gamma _\mu
-iA_\mu )+{M}+\tilde V(r)\right\} \Psi ~=~0
\end{equation}
where $\tilde \gamma ^\mu $ are the curved gamma matrices satisfying the
relation, $\left\{ {\tilde \gamma }^\mu ,\tilde \gamma ^\nu \right\}_{+
}=2g^{\mu \nu }$, and $\Gamma _\mu $ are the spin connections \cite{Brill}.
with $\tilde V(r)$ as the scalar 1/r field
\begin{equation}
\tilde V(r)=-\frac{\alpha ^{\prime }}r
\end{equation}
where $\alpha ^{\prime }$ is a constant, and the vector potential $A_\mu $
reads
\begin{equation}
A_\mu =A_\mu ^{({\rm mon})}+A_\mu ^{({\rm Coul})}+A_\mu ^{({\rm AB})}
\end{equation}
where the components of the Coulomb potential $A_\mu ^{({\rm Coul})}$ take
the form
\begin{equation}
A_0^{({\rm Coul})}=V(r)=-\frac \alpha r,\ A_i^{({\rm Coul})}=0,\ i=1,2,3
\end{equation}
the Aharonov Bohm potential $A_\mu ^{({\rm AB})}$ reads
\begin{equation}
A^{({\rm AB})}=\frac F{r \sin \vartheta }\hat e_\varphi
\end{equation}
and the Dirac monopole field $A_\mu ^{({\rm mon})}$ is
\begin{equation}
A^{({\rm mon})}=g\frac{(1-\cos \vartheta )}{r\sin \vartheta }\hat e_\varphi
\end{equation}
where, following the Dirac \cite{Dirac} prescription for quantizing the
magnetic charge, $g$ takes integer or half integer values
\begin{equation}
\label{g}g=\frac j2,\ j=0,\ \pm 1,\ \pm 2....
\end{equation}

If we choose to work in the fixed Cartesian gauge, the spinor connections
are zero and the $\tilde \gamma $ matrices take the form
$$
\tilde \gamma ^0={\gamma }^0=\bar \gamma ^0,\ \tilde \gamma ^1=\left[
(\gamma ^1\cos \varphi +\gamma ^2\sin \varphi )\sin \vartheta +\gamma ^3\cos
\vartheta \right] =\bar \gamma ^1,
$$
\begin{equation}
\label{carte}\tilde \gamma ^2=\frac 1r\left[ (\gamma ^1\cos \varphi +\gamma
^2\sin \varphi )\cos \vartheta -\gamma ^3\sin \vartheta \right] =\frac{\bar
\gamma ^2}r,\quad
\end{equation}
$$
\tilde \gamma ^3=\frac 1{r\sin \vartheta }(-\gamma ^1\sin \varphi +\gamma
^2\cos \varphi )=\frac{\bar \gamma ^3}{r\sin \vartheta }
$$
where $\gamma^{\alpha}$ are the standard Minkowski gamma matrices,
and the Dirac equation in the fixed Cartesian tetrad frame (\ref{carte})
takes the form
\begin{equation}
\label{car}
\left\{ \bar \gamma ^0(\partial _t+iV(r))+\bar \gamma ^1\partial _r+\frac{%
\bar \gamma ^2}r\partial _\vartheta +\frac{\bar \gamma ^3}{r\sin \vartheta }%
(\partial _\varphi -iF-i(1-\cos \vartheta )g)+{M+}\tilde V(r)\right\}
\Psi_{Cart}~=~0
\end{equation}
where we have introduced the spinor $\Psi_{Cart}$, solution of the Dirac
equation
(\ref{car}) in the fixed tetrad gauge.
In order to separate variables in the Dirac equation, we are going to work
in the diagonal tetrad gauge where the gamma matrices $\tilde \gamma _d$
take the form
\begin{equation}
\label{diag}\tilde \gamma _d^0=\gamma ^0,\ \tilde \gamma _d^1=\gamma ^1,\
\tilde \gamma _d^2=\frac 1r\gamma ^2,\ \tilde \gamma _d^3=\frac 1{r\sin
\vartheta }\gamma ^3
\end{equation}

Since the curvilinear matrices $\tilde \gamma ^\mu $ and $\tilde
\gamma _d$ satisfy the same anticommutation relations, they are related by a
similarity transformation, unique up to a factor. In the present case we
choose this factor in order to eliminate the spin connections in the
resulting Dirac equation. The transformation S can be written as \cite
{victor}

\begin{equation}
\label{vest}S=\frac 1{r(\sin \vartheta )^{1/2}}\exp (-\frac \varphi 2\gamma
^1\gamma ^2)\exp (-\frac \vartheta 2\gamma ^3\gamma ^1){\sf a=S}_0{\sf a}
\end{equation}
where ${\sf a}$ is a constant non singular matrix given by,
${\sf a}=\frac 12(\gamma ^1\gamma ^2-\gamma ^1\gamma ^3+\gamma ^2\gamma ^3+I)$
which applied on the gamma's acts as follows
\begin{equation}
\label{a'}{\sf a}\gamma ^1{\sf a}^{-1}=\gamma ^3,\ {\sf a}\gamma ^2{\sf a}%
^{-1}=\gamma ^1,\ {\sf a}\gamma ^3{\sf a}^{-1}=\gamma ^2,
\end{equation}
the transformation S acts on the curvilinear $\tilde \gamma $ matrices,
reducing them to the rotating diagonal gauge as follows
\begin{equation}
S^{-1}\tilde \gamma ^\mu S=g^{\mu \mu }\gamma ^\mu =\tilde \gamma _d^\mu
\quad {\rm (no\ summation)}
\end{equation}
then, the Dirac equation in spherical coordinates, with the radial potential
$V(r)$, in the local rotating frame reads

\begin{equation}
\label{uno}\left\{ \gamma ^0(\partial _t+iV(r))+\gamma ^1\partial _r++\frac{%
\gamma ^2}r\partial _\vartheta +\frac{\gamma ^3}{r\sin \vartheta }(\partial
_\varphi -iF-i(1-\cos \vartheta )g)+{M}+\tilde V(r)\right\} \Psi_{rot} =0
\end{equation}
\noindent where we have introduced the spinor $\Psi_{rot}$, related to
$\Psi_{Cart}$ by the expression

\begin{equation}
\label{16}\Psi_{Cart} =S\Psi_{rot} =S_0{\sf a}\Psi_{rot}
\end{equation}
\noindent and $\gamma ^\mu $ are the standard Dirac flat matrices.

\noindent Applying the algebraic method of separation of variables \cite
{alg1,alg2}, it is possible to write eq. (\ref{uno}) as a sum of two first
order linear differential operators $\hat K_1$ , $\hat K_2$ satisfying the
relation

\begin{equation}
\label{17}\left[ \hat K_1,\hat K_2\right] =0,\hspace{2cm}\left\{ \hat K_1+{%
\hat K}_2\right\} \Phi =0
\end{equation}
\begin{equation}
\label{18}{\hat K}_2\Phi =k\Phi =-{\hat K}_1\Phi
\end{equation}
then, if we separate the time and radial dependence from the angular one, we
obtain
\begin{equation}
\label{dos}{\hat K}_2\Phi =-i\left[ \gamma ^2\partial _\vartheta +\frac{%
\gamma ^3}{\sin \vartheta }(\partial _\varphi -iF-i(1-\cos \vartheta
)g)\right] \gamma ^0\gamma ^1\Phi =k\Phi
\end{equation}
\begin{equation}
\label{quatre}{\hat K}_1\Phi =-ir\left[ \gamma ^0\partial _t+\gamma
^1\partial _r+{M}+i\gamma ^0V(r)+\tilde V(r)\right] \gamma ^0\gamma
^1\Phi =-k\Phi
\end{equation}
with
\begin{equation}
\label{ici}\Psi_{rot} =\gamma ^0\gamma ^1\Phi
\end{equation}
where the operator \^K$_1$ as well as ${\hat K}_2$ have been chosen to be
Hermitean, therefore the constant of separation $k$ appearing in (\ref{dos})
and (\ref{quatre}) is real. Notice that if we drop out the Aharonov-Bohm and
the magnetic charge contributions in eq. (\ref{dos}) we obtain the Brill and
Wheeler \cite{Brill} angular momentum \^K, and separation constant $k$ takes
integer values.

Now we proceed to decouple the equation (\ref{dos}) governing the angular
dependence of the Dirac spinor In order to simplify the resulting equations,
we choose to work with the auxiliary spinor $\bar \Phi $ related to $\Phi $
as follows
\begin{equation}
\bar \Phi ={\sf a}\Phi
\end{equation}
Since the operators $\hat K_2$ and $\hat K_1$ commute with the projection of
the angular momentum $-i\partial_{\phi}$, with eigenvalues $m$, we have that
equation (\ref{dos}) takes the form
\begin{equation}
\label{cinq}\left[ \gamma ^1\partial _\vartheta -i\frac{\gamma ^2}{\sin
\vartheta }(m +F+(1-\cos \vartheta )gt)\right] \gamma
^0\gamma ^3\bar \Phi =-ik\bar \Phi
\end{equation}

In order to reduce the equation (\ref{cinq}) to a system of ordinary
differential equations, we choose to work in the following representation
for the gamma matrices, \cite{Jauch}
\begin{equation}
\label{tres}\gamma ^i=\left(
\begin{array}{cc}
0 & \sigma ^i \\
\sigma ^i & 0
\end{array}
\right) ,\gamma ^0=\left(
\begin{array}{cc}
-i & 0 \\
0 & i
\end{array}
\right) ,
\end{equation}

Then, substituting (\ref{tres}) into (\ref{cinq}) we obtain,
\begin{equation}
\label{ang1}\left[ \sigma ^2\partial _\vartheta -i\frac{\sigma ^1}{\sin
\vartheta }(m+F+(1-\cos \vartheta )g)\right] \bar \Phi _1=-ik\bar \Phi _1
\end{equation}
\begin{equation}
\label{ang2}\left[ -\sigma ^2\partial _\vartheta +i\frac{\sigma ^1}{\sin
\vartheta }(m+F+(1-\cos \vartheta )g)\right] \bar \Phi _2=-ik\bar \Phi _2
\end{equation}
with
\begin{equation}
\bar \Phi =e^{im\phi}\left(
\begin{array}{c}
\bar \Phi _1 \\
\bar \Phi _2
\end{array}
\right)
\end{equation}

\section{Solution of the angular equations}

In this section we are going to solve the systems of equations (\ref{ang1})
and (\ref{ang2}). It is not difficult to see that the spinor $\bar \Phi _2$
can be written as $G(r,t)\sigma ^3\bar \Phi _1$, where $G(r,t)$ is an arbitrary
function, this property allows us to
consider only the system (\ref{ang1}). Using the standard Pauli matrices ,
we have that (\ref{ang1}) reduces to
\begin{equation}
\label{tsi1}k\sin \vartheta \xi _1-(m-F-(1-\cos \vartheta )g)\xi _2+\sin
\vartheta \frac{d\xi _2}{d\vartheta }=0
\end{equation}
\begin{equation}
\label{tsi2}-k\sin \vartheta \xi _2+(m-F-(1-\cos \vartheta ))\xi _1+\sin
\vartheta \frac{d\xi _1}{d\vartheta }=0
\end{equation}
with
\begin{equation}
\label{Fi1}\bar \Phi _1=Qe^{im\phi}\left(
\begin{array}{c}
\xi _1 \\
\xi _2
\end{array}
\right)
\end{equation}
where Q is a function depending on the variables $t$, $r$, to be
determined after a complete separation of variables.
In order to solve the coupled system of equations (\ref{tsi1})-(\ref{tsi2})
we make the following ansatz,
\begin{equation}
\label{3'}\quad \xi _1=(\sin \frac \vartheta 2)^a(\cos \frac \vartheta
2)^bf(\vartheta ),\ \xi _2=(\sin \frac \vartheta 2)^c(\cos \frac \vartheta
2)^dq(\vartheta )
\end{equation}
where a, b, c, and d are constants to be fixed in order to obtain solutions
of the governing equations for $f(\vartheta )$ and $q(\vartheta )$ in terms
of orthogonal special functions. Substituting (\ref{3'}) into (\ref{tsi1})
and (\ref{tsi2}) we obtain,
\begin{equation}
\label{6'}-kq(x)-(\frac m2-\frac F2+\frac a2+\frac 12)f(x)+(1-x)\frac{df(x)}{%
dx}=0
\end{equation}
\begin{equation}
\label{7'}kf(x)+(\frac d2-g-\frac m2-\frac F2)q(x)+(1+x)\frac{dq(x)}{dx}=0
\end{equation}
where we have made the change of variable $x=\cos \vartheta ,$ and we have
simplified the resulting equations (\ref{6'}) and (\ref{7'}) by imposing $%
c=m-F,$ and $b=m-F-2g.$ If we set $a=m-F+1,$ and $d=m-F+1-2g,$and make the
change of variables $u=(1-x)/2$, we obtain that the coupled system of
equations (\ref{6'})-(\ref{7'}) takes the form
\begin{equation}
\label{8'}-kq(u)-(m-F+\frac 12)f(u)-u\frac{df(u)}{du}=0
\end{equation}
\begin{equation}
\label{9'}kf(u)+(m-F-2g+\frac 12)q(u)+(u-1)\frac{dq(u)}{du}=0
\end{equation}
from (\ref{8'}) and (\ref{9'}) we obtain%
$$
u(1-u)\frac{d^2q(u)}{du^2}+[(m-F+\frac 12)-2(m-F-g+\frac 12)u]\frac{dq(u)}{du%
}+
$$
\begin{equation}
\label{2q}+[k^2-(m-F+\frac 12)(m-F-2g+\frac 12)]q(u)=0
\end{equation}
$$
u(1-u)\frac{d^2f(u)}{du^2}+[(m-F+\frac 32)-2(m-F-g+1)u]\frac{df(u)}{du}+
$$
\begin{equation}
\label{2f}+[k^2-(m-F+\frac 12)(m-F-2g+\frac 12)]f(u)=0
\end{equation}
the solution of the equation (\ref{2q}) can be expressed in terms of the
hypergeometric function $F(a,b;c;u)$ \cite{Gradshteyn} as follows
\begin{equation}
\label{q}q(u)=c_1F(a,b;c;u)
\end{equation}
where c$_1$ is a constant, and $a,b,$ and $c$ are
\begin{equation}
a=m-F-g+\frac 12-\sqrt{k^2+g^2}
\end{equation}
\begin{equation}
b=m-F-g+\frac 12+\sqrt{k^2+g^2}
\end{equation}
\begin{equation}
c=m-F+\frac 12
\end{equation}
then, with the help of (\ref{9'}) and (\ref{q}) we find that $f(u)$ reads
\begin{equation}
\label{f}f(u)=c_1\frac k{m-F+\frac 12}F(a,b;c+1;u)
\end{equation}
Since we are looking for normalizable solutions according to the product
\begin{equation}
\label{int}2\pi \int_0^{2\pi }\Phi _k^{\mu \dagger }\Phi _{k^{\prime }}^\mu
\ d\vartheta =\delta _{kk^{\prime }}
\end{equation}
we have that the series associated with the hypergeometric function $%
F(a,b;c;u)$ should be truncated reducing it to polynomials. This one is
possible if $a=-n$ or $b=-n$ in (\ref{q}) where $n$ is a no negative integer
value. Then using the relation between the Jacobi Polynomials and the
functions $F(a,b;c;x)$
\begin{equation}
\label{Jac}P_n^{(\alpha ,\beta )}(x)=\frac{\Gamma (n+\alpha +1)}{n!\Gamma
(\alpha +1)}F(-n,n+\alpha +\beta +1;\alpha +1;\frac{1-x}2)
\end{equation}
we have that (\ref{f}) and (\ref{q}) reduce respectively to
\begin{equation}
f(x)=c^{\prime }P_n^{(m-F+1/2,m-F-2g-1/2)}(x)
\end{equation}
\begin{equation}
q(x)=c^{\prime }\frac{g+\sqrt{k^2+g^2}}kP_n^{(m-F-1/2,m-F-2g+1/2)}(x)
\end{equation}
where $c^{\prime }$ is a constant, and $n$ reads
\begin{equation}
\label{a:}n=-m+F+g-\frac 12+\sqrt{k^2+g^2}
\end{equation}
Then, the components $\xi _1$ and $\xi _2$ of the spinor $\bar \Phi $ (\ref
{Fi1}) can be written as
\begin{equation}
\label{T1}\xi _1=c^{\prime }(\sin \frac \vartheta 2)^{m-F+1}(\cos \frac
\vartheta 2)^{m-F-2g}P_n^{(m-F+1/2,m-F-2g-1/2)}(\cos \vartheta )
\end{equation}
\begin{equation}
\label{T2}\xi _2=c^{\prime }\frac{g+\sqrt{k^2+g^2}}k(\sin \frac \vartheta
2)^{m-F}(\cos \frac \vartheta 2)^{m-F-2g+1}P_n^{(m-F-1/2,m-F-2g+1/2)}(\cos
\vartheta )
\end{equation}
Notice that the orthogonality relation for the Jacobi Polynomials%
\begin{eqnarray}
\label{norme}
\int^{1}_{-1} (1-x)^\alpha (1+x)^\beta P_n^{(\alpha ,\beta )}(x)P_m^{(\alpha
,\beta
)}(x)dx= \nonumber  \\
\frac{2^{\alpha +\beta +1}}{2n+\alpha +\beta +1}\frac{\Gamma
(n+\alpha +1)\Gamma (n+\beta +1)}{\Gamma (n+1)\Gamma (n+\alpha +\beta +1)}%
\delta _{mn}
\end{eqnarray}
imposes some restrictions on the values of $m,$ $F$ and $g$ in (\ref{T1})
and (\ref{T2}) In fact, from (\ref{norme}) we have that $\alpha >-1,$ $\beta
>-1$ \cite{Magnus} and consequently we have that
\begin{equation}
\label{a}m-F+\frac 12>0,\ m-F-2g+1/2>0
\end{equation}
affords the required condition of orthogonality. However, some values
considered in the inequalities given by (\ref{a}) fail in fulfilling a
condition which should be also took into account in selecting the possible
values of $m,$ $F$ and $g.$ : All the expectation values associated with the
separating operators should exist. In fact, if we consider the operator \^K$%
_2$ defined by the relation (\ref{dos}) we have that the integrals $\int
d\vartheta \xi _1\partial _\vartheta \xi _1$ as well as $\int d\vartheta \xi
_2\partial _\vartheta \xi _2$ should be finite. This one imposes some
restrictions on $m,$ $F$ and $g$. Looking at the convergency when $\vartheta
\rightarrow 0,$ and $\vartheta \rightarrow \pi $ we obtain the relations
\begin{equation}
\label{a"}m-F>0,\ m-F-2g>0
\end{equation}
which are weaker than those given in (\ref{a})

In order to be able to consider other relations among the parameters $m,$ $F$
and $g$ different from (\ref{a:}) we are going to consider a second solution
of the equation differential equation (\ref{2q})
\begin{equation}
q(u)=u^{1-c}(1-u)^{c-a-b}F(1-a,1-b;2-c;u)
\end{equation}
then, using the recurrence relations for the hypergeometric functions, we
find that a solution for the system (\ref{8'})-(\ref{9'}) reads
\begin{equation}
\label{efe}f(u)=c_2u^{-1/2-m+F}(1-u)^{-m+F+2g+1/2}F(1-a,1-b;\frac 12-m+F;u)
\end{equation}
\begin{equation}
\label{cu}q(u)=c_2\frac k{\frac
12-m+F}u^{-1/2-m+F}(1-u)^{-m+F+2g+1/2}F(1-a,1-b;\frac 32-m+F;u)
\end{equation}
where $c_2$ is a constant. From (\ref{efe}) and (\ref{cu}) we obtain that $%
\xi _1$ and $\xi _2$ take the form
\begin{equation}
\label{T3}\xi _1=c(\sin \frac \vartheta 2)^{-m+F}(\cos \frac \vartheta
2)^{-m+F+2g+1}P_n^{(-1/2-m+F,2g-m+F+1/2)}(\cos \vartheta )
\end{equation}
\begin{equation}
\label{T4}\xi _2=c\frac k{\sqrt{k^2+g^2}-g}(\sin \frac \vartheta
2)^{1-m+F}(\cos \frac \vartheta 2)^{-m+F+2g}P_n^{(1/2-m+F,2g-m+F-1/2)}(\cos
\vartheta )
\end{equation}
with c as a constant of normalization, and in the present case $n$ is
\begin{equation}
\label{b:}n=m-F-g-\frac 12+\sqrt{k^2+g^2}
\end{equation}
Following the same reasoning used for deriving (\ref{a"}), we have that the
expressions for $\xi _1$ and $\xi _2$ given by (\ref{T3}) and (\ref{T4}) are
valid when
\begin{equation}
\label{b}-m+F>0,\ 2g-m+F>0
\end{equation}
A third possible solution of the equation (\ref{2q}) can be written as

\begin{equation}
\label{cu1}q(u)=(1-u)^{c-a-b}F(c-a,c-b;c;u)
\end{equation}
then, substituting (\ref{cu1}) into (\ref{9'}) we find that $f(u)$ and $q(u)$
take the form:
\begin{equation}
f(u)=c_3(1-u)^{-m+F+2g+1/2}F(1+g+\sqrt{k^2+g^2},1+g-\sqrt{k^2+g^2};m-F+\frac
32;u)
\end{equation}
\begin{equation}
q(u)=-c_3\frac{m-F+1/2}k(1-u)^{-m+F+2g-1/2}F(g+\sqrt{k^2+g^2},g-\sqrt{k^2+g^2%
};m-F+\frac 12;u)
\end{equation}
where c$_3$ is a constant. Using the relation between the hypergeometric
function $F(a,b;c;x)$ and the Jacobi Polynomials P$_n^{(\alpha ,\beta )}(x)$
given by eq. (\ref{Jac}) we arrive at
\begin{equation}
\label{T5}\xi _1=c(\sin \frac \vartheta 2)^{m-F+1}(\cos \frac \vartheta
2)^{-m+F+2g+1}P_n^{(m-F+1/2,-m+F+2g+1/2)}(\cos \vartheta )
\end{equation}
\begin{equation}
\label{T6}\xi _2=-c\frac{n+1}k(\sin \frac \vartheta 2)^{m-F}(\cos \frac
\vartheta 2)^{-m+F+2g}P_{n+1}^{(m-F-1/2,-m+F+2g-1/2)}(\cos \vartheta )
\end{equation}
where $c$ is a constant of normalization and $n$ is given by
\begin{equation}
\label{N}n=\sqrt{k^2+g^2}-g
\end{equation}
In this case, we have to impose the following restrictions on the values of $%
m,$ $F$ and $g.$%
\begin{equation}
\label{c}m-F>0,\ -m+F+2g>0
\end{equation}
Finally, considering as solution of the equation (\ref{2q}) the expression
\begin{equation}
q(u)=u^{1-c}F(a-c+1,b-c+1;2-c;u)
\end{equation}
then, in the present case the functions $f(u)$ and $q(u)$ read
\begin{equation}
f(u)=u^{-1/2-m+F}F(-g-\sqrt{k^2+g^2},-g+\sqrt{k^2+g^2};\frac 12-m+F;u)
\end{equation}
\begin{equation}
q(u)=\frac k{\frac 12-m+F}u^{1/2-m+F}F(-g-\sqrt{k^2+g^2}+1,-g+\sqrt{k^2+g^2}%
+1;\frac 32-m+F;u)
\end{equation}
\begin{equation}
\label{T7}\xi _1=c_4(\sin \frac \vartheta 2)^{-m+F}(\cos \frac \vartheta
2)^{m-F-2g}P_n^{(-1/2-m+F,1/2+m-F-2g)}(\cos \vartheta )
\end{equation}
\begin{equation}
\label{T8}\xi _2=c_4\frac kn(\sin \frac \vartheta 2)^{1-m+F}(\cos \frac
\vartheta 2)^{m-F-2g+1}P_{n-1}^{(1/2-m+F,-1/2+m-F-2g)}(\cos \vartheta )
\end{equation}
with $n$ given by
\begin{equation}
\label{N2}n=\sqrt{k^2+g^2}+g
\end{equation}
and c$_4$ is a constant of normalization. The solutions (\ref{T7}) and (\ref
{T8}) are well behaved according to the Dirac inner product as well as to
the expectation value of the operator of angular momentum (\ref{dos}) if
\begin{equation}
\label{d}-m+F>0,\ m-F-2g>0
\end{equation}
then, we have that the results (\ref{N}) and (\ref{N2}) can be gathered as
follows
\begin{equation}
\label{N3}n=\sqrt{k^2+g^2}-\mid g\mid
\end{equation}
when the condition on $m$, $F$ and $g$
\begin{equation}
\label{N4}F+g+\mid g\mid >m>g-\mid g\mid +F
\end{equation}
is satisfied.

Regarding the eigenvalues $m$ of the projection of the angular momentum
operator $-i\partial _\varphi $ we have that since the transformation (\ref
{vest}), relating the Dirac spinors $\Psi_{rot}$ and $\Psi_{Cart}$ in the local
(rotating) and the Cartesian tetrad frames transforms after a rotation as
follows
\begin{equation}
S_z(\varphi +2\pi )=-S_z(\varphi )
\end{equation}
and the spinor $\Psi_{Cart} $ is single valued, then we obtain
\begin{equation}
\Psi_{rot}(\varphi +2\pi )=-\Psi_{rot} (\varphi )
\end{equation}
and therefore $m$ takes half integer values
\begin{equation}
\label{m}m={\tt N}+\frac 12,\ {\tt N}=0,\ \pm 1,\ \pm 2...
\end{equation}

\section{Solution of the radial equation}

Now, we are going to solve the system of equation (\ref{quatre}), governing
the radial dependence of the spinor $\Psi_{rot}$ solution of the Dirac
equation.
This equation can be written in the form,
\begin{equation}
\label{coco}
\left( -\gamma ^3\partial _t+\gamma ^0\partial _r+(M+\tilde V(r))\gamma
^0\gamma
^3-i\gamma ^3V(r)+i\frac kr\right) \bar \Phi =0
\end{equation}
Using the representation for the gamma matrices given by ($\ref{tres}$), and
the fact that eq. (\ref{coco}) commutes with the energy operator $\i\partial_t$
with eigenvalues $E$, we obtain the following system of equations
\begin{equation}
\label{lap}\left( d_r+\frac kr\right) \bar \Phi _1-\sigma ^3(E-V(r)-M-\tilde
V(r))\bar
\Phi _2=0
\end{equation}
\begin{equation}
\label{las}\left( -d_r+\frac kr\right) \bar \Phi _2-\sigma ^3(E-V(r)+M+\tilde
V(r))\bar
\Phi _1=0
\end{equation}
{}From (\ref{ang1})-(\ref{ang2}),(\ref{lap})-(\ref{las}) and the fact that
the Dirac equation (\ref{uno}) commutes with $-i\partial_{\phi}$ and
$i\partial_{t}$, we have that the spinor $\bar \Phi $ can be written as
follows
\begin{equation}
\label{sol}\bar \Phi
=c_0e^{i(m\varphi -Et)}\left(
\begin{array}{c}
\xi _1
{\sf A}(r) \\ \xi _2
{\sf A}(r) \\ c\xi _1
{\sf B}(r) \\ -c\xi _2{\sf B}(r)
\end{array}
\right)
\end{equation}
where $c$ is a constant, and ${\sf A}(r)$ and ${\sf B}(r)$ satisfy the system
of
equations
\begin{equation}
\label{pa}\left( d_r+\frac kr\right) {\sf A}(r)-(E-V(r)-\tilde V(r)-{M})%
{\sf B}(r)=0
\end{equation}
\begin{equation}
\label{pe}\left( -d_r+\frac kr\right) {\sf B}(r)-(E-V(r)+\tilde V(r)+{M})%
{\sf A}(r)=0
\end{equation}
notice that in this way we have fixed the values of the functions
$Q(r,t)=e^{-iEt}{\sf A(r)}$, and $G={\sf B(r)}/{\sf A(r)}$ appearing during the
process of separation of variables.
Substituting into (\ref{pa}) and (\ref{pe}) the form of the scalar and the
Coulomb potentials we get
\begin{equation}
\label{R1}\left( \frac d{dr}+\frac kr\right) {\sf A}(r)-\left( E-{M}%
+\frac 1r(\alpha +\alpha ^{\prime })\right) {\sf B}(r)=0
\end{equation}
\begin{equation}
\label{R2}\left( \frac d{dr}-\frac kr\right) {\sf B}(r)+\left( E+{M}%
-\frac 1r(\alpha -\alpha ^{\prime })\right) {\sf A}(r)=0
\end{equation}
Introducing the notation
\begin{equation}
\label{val}A=E+{M},\ B={M}-E,\ \hat \alpha =\alpha +\alpha ^{\prime
},\ \beta =\alpha ^{\prime }-\alpha
\end{equation}
and the new variable $\rho ,$ related to $r$ as follows
\begin{equation}
\rho =Dr=\sqrt{{M}^2-E^2}r=\sqrt{AB}r
\end{equation}
we have that the system of equations (\ref{R1})-(\ref{R2}) reduces to
\begin{equation}
\label{R3}\left( \frac d{d\rho }+\frac k\rho \right) {\sf A}(r)-\left( \frac{%
-B}D+\frac{\hat \alpha }\rho \right) {\sf B}(r)=0
\end{equation}
\begin{equation}
\label{R4}\left( \frac d{d\rho }-\frac k\rho \right) {\sf B}(r)+\left( \frac
AD-\frac{\hat \beta }\rho \right) {\sf A}(r)=0
\end{equation}
We shall look for solutions of the system (\ref{R3})-(\ref{R4}) in the form
of power series \cite{alg3,Davydov}
\begin{equation}
\label{R5}{\sf A}(\rho )=e^{-\rho }\sum_{\nu =0}^\infty \rho ^{s+\nu }a_\nu
\end{equation}
\begin{equation}
\label{R6}B(\rho )=e^{-\rho }\sum_{\nu =0}^\infty \rho ^{s+\nu }b_\nu
\end{equation}
Substituting (\ref{R5})-(\ref{R6}) into (\ref{R3})-(\ref{R4}) we find
\begin{equation}
\label{R7}(s+k)a_0-\hat \alpha b_0=0
\end{equation}
\begin{equation}
\label{R8}(s-k)b_0-\hat \beta a_0=0
\end{equation}
and
\begin{equation}
\label{R9}\left[ (s+\nu )+k\right] a_\upsilon -\hat \alpha b_\nu +\frac
BDb_{\nu -1}-a_{\nu -1}=0
\end{equation}
\begin{equation}
\label{R10}\left[ (s+\nu )-k\right] b_\nu -\hat \beta a_\nu +\frac ADa_{\nu
-1}-b_{\nu -1}=0
\end{equation}
{}From (\ref{R7})-(\ref{R8}) it follows that
\begin{equation}
\label{R11}s=\sqrt{k^2-\alpha ^2+\alpha ^{\prime 2}}
\end{equation}
where we have dropped out (\ref{R11}) the negative root because we are
looking for wavefunctions regular at the origin of coordinates. From (\ref
{R9})-(\ref{R10}) we have that
\begin{equation}
\label{R12}\left\{ \left[ (s+\nu )+k\right] \sqrt{\frac AB}-\hat \beta
\right\} a_\upsilon =\left\{ \sqrt{\frac AB}\hat \alpha -\left[ (s+\nu
)-k\right] \right\} b_\upsilon
\end{equation}
The series (\ref{R5}) and (\ref{R6}) will have a good behavior at infinity
if they terminate for a finite value N. Putting $a_{N+1}=b_{N+1}=0$ in (\ref
{R9})-(\ref{R10}) with $a_N\neq 0$ and $b_N\neq 0,$ we arrive at
\begin{equation}
\label{R13}\frac{b_N}{a_N}=\frac AD
\end{equation}
Substituting (\ref{R13}) into (\ref{R12}), and taking into account (\ref{val}%
) we get
\begin{equation}
\label{R14}(s+N)\sqrt{{M}^2-E^2}=E\alpha +\alpha ^{\prime }{M}
\end{equation}
where $s$ is given by (\ref{R11})

Finally, we obtain the energy spectrum:
\begin{equation}
\label{R15}E={M}\left\{ -\frac{\alpha \alpha ^{\prime }}{(s+N)^2+\alpha
^2}\pm \sqrt{\left( \frac{\alpha \alpha ^{\prime }}{(s+N)^2+\alpha ^2}%
\right) ^2-\frac{\alpha ^{\prime 2}-(s+N)^2}{\left[ (s+N)^2+\alpha ^2\right]
^2}}\right\}
\end{equation}
Here two particular cases could be considered: a) $\alpha ^{\prime }=0$
which corresponds to the Coulomb potential. In this case the energy spectrum
reduces to:
\begin{equation}
\label{R16}E=M\left[ 1+\frac{\alpha ^2}{(s+N)^2}\right] ^{-1/2},\ s=\sqrt{%
k^2-\alpha ^2}
\end{equation}
where the negative root has been dropped out because it is not compatible
with the relation (\ref{R14}) A second possibility is given by b) $\alpha
=0, $ which is the scalar $V\prime (r)=-\alpha ^{\prime }/r$ potential. In
this case the energy spectrum takes the form
\begin{equation}
\label{R17}E=\pm {M}\left[ 1-\frac{\alpha ^{\prime 2}}{(s+N)^2}\right]
^{1/2},\ \ s=\sqrt{k^2+\alpha ^{\prime 2}}
\end{equation}
notice that in the present case states with negative energy are possible,
here we do not have critical behavior like in the Coulomb case.

\section{Discussion of the results}

In this section we are going to discuss the influence of the Aharonov-Bohm
potential and the Dirac magnetic monopole charge on the energy spectrum.
Here we have mention that the Aharonov-Bohm as well as the magnetic monopole
contributions are present in the expression (\ref{R15}) via the factor $s$
given by (\ref{R11}), since the explicit form of $k$ depends on the relation
among $m,F$ and $g.$ In fact, we have that when the inequalities (\ref{c})
or (\ref{d}) are valid, the expression for $s$ takes the form
\begin{equation}
\label{D1}s=\sqrt{(n+\mid g\mid )^2-g^2-\alpha ^2+\alpha ^{\prime 2}}
\end{equation}
and no contribution of the Aharonov Bohm potential is observed in (\ref{R15}%
). The values of $m$ for which (\ref{D1}) takes place are given by the
expression (\ref{N4})

It is worth mentioning that when the magnetic monopole contribution is
absent, the inequalities given by the expression (\ref{N4})
never take place, and consequently, the expression for $s$ given by eq. (\ref
{D1}) is not applicable. In this case the energy spectrum can be computed by
substituting the expression
\begin{equation}
\label{D5}s=\sqrt{(n+\mid m-F\mid +1/2)^2-\alpha ^2+\alpha ^{\prime 2}}
\end{equation}
into (\ref{R15}). Analogously, the energy spectrum  when $F\neq 0$ and g$%
\neq 0$ can be obtained after substituting into (\ref{R15}) the following
value of $s$
\begin{equation}
\label{D6}s=\sqrt{(n+m-F-g+1/2)^2-g^2-\alpha ^2+\alpha ^{\prime 2}}
\end{equation}
when $m-F>0$ and $m-f-2g>0.$ Otherwise, when  $m-F<0$ and $m-F-2g<0$, the
value of $s$ to substitute into (\ref{R15}) reads
\begin{equation}
\label{D7}s=\sqrt{(n-m+F+g+1/2)^2-g^2-\alpha ^2+\alpha ^{\prime 2}}
\end{equation}
Perhaps the most interesting and puzzling result of the present paper is the
non dependence of the energy spectrum on the Aharanov-Bohm potential for a
range of values given by the inequality (\ref{N4}). Despite this phenomenon
was already pointed by Hoang {\it et al} \cite{Komarov} there are some
discrepancies between their results and those ones present in this paper.
Basically the problems lie on the criteria for establishing the boundary
conditions and the normalizability of the wave functions. It is worth
mentioning that since the  Aharonov-Bohm contribution $F$ can take non
integer values, then the parameters $\alpha $ and $\beta $ in the Jacobi
Polynomials P$_n^{(\alpha ,\beta )}(x)$ can be negative provided that $%
\alpha >-1,$ and $\beta >-1.$ Obviously if the Aharonov Bohm potential is
absent, from the results presented in Sec. 3 we have that $\alpha \geq 0$
and $\beta \geq 0$ (or in the case when $F$ is an integer).  Hoang {\it et
al }consider that $\alpha $ and $\beta $ are always positive restricting in
the way the range of validity of the solutions. A second point to remark is
that we not only impose the normalizability of the wave functions but also
the existence of the expectation value of the angular momentum operator,
which is equivalent to say that $\int \Phi ^{\dagger }\ \hat K$ $\Phi
d\vartheta d\varphi <\infty ,$ in this way we have to impose that the spinor
components $\xi _1$and $\xi _2$ presented in Sec. 3 should satisfy $\int \xi
_{1,2}\partial _\vartheta \xi _{1,2}<\infty $ . Regarding the assertion made
in \cite{Komarov} about the existence of quantum states forbidden for the
Dirac particle in the presence of the Coulomb plus the Dirac monopole
potentials, we have that the boundary conditions imposed on the wave
function in the present article avoid such a anomalous behavior and
consequently the spinor $\Phi $ is well defined for any value of the
parameters $m,$ $F$ and $g.$

\acknowledgments

\noindent The author wishes to express his indebtedness to the Centre de
Physique Th\'eorique for the suitable conditions of work. Also the author
wishes to acknowledge to the CONICIT of Venezuela and to the Fundaci\'on
Polar for financial support.

\end{document}